\documentclass[12pt]{article}
\renewcommand{\c} {\cite}
\def\be{\begin{equation}}
\def\ee{\end{equation}}
\def\bea{\begin{eqnarray}}
\def\eea{\end{eqnarray}}
\newcommand{\nn}{\nonumber}

\newcommand{\Tr}{{\rm Tr}}
\def\bal{\begin{eqnarray}}
\def\eal{\end{eqnarray}}
\def\lesssim{\stackrel{<}{\sim}}
\def\gtrsim{\stackrel{>}{\sim}}
\begin{document}
\begin{titlepage}
\setcounter{page}{0}
\rightline{}
\vspace{2cm}
\begin{center}
{\LARGE\bf On the polarization of the nucleon sea \\
in the meson cloud model \\}
\vspace*{1cm}
{\large K.G. Boreskov, A.B. Kaidalov} \\
\vspace*{3mm}
{\em ITEP, B.Cheremushkinskaya 25 \\
117259 Moscow, Russia}\\
\end{center}
\vspace{5mm}
\centerline{{{\bf Abstract}}}

It is pointed out that the meson cloud model predicts a
substantial polarization of the sea quarks of the nucleon due to
interference of $\pi$ and $\rho$ exchanges. This polarization is
strongly flavour dependent. The model gives an explanation of a
strong increase in the structure function $g_1$ with isospin
$I=1$ at small $x$.

\end{titlepage}
\newpage
\renewcommand{\thefootnote}{\arabic{footnote}}
\setcounter{footnote}{0}
\section*{1. Introduction}
\indent

Experimental studies of nucleon structure functions have revealed
several interesting and unexpected features. Investigation of the
Gottfried sum rule by the NMC collaboration \c{NMC} lead to the
conclusion that $\bar{u}$ and $\bar{d}$ quarks distributions in
the proton are different. This result has been confirmed by
measurements of the Drell--Yan process \c{DY}, \c{E866}. The
difference between $\bar{u}$ and $\bar{d}$ distributions is
measured now as a function of $x$ \c{E866} and is concentrated in
the region of $x\lesssim 0.2$.
This phenomenon has a natural explanation \c{ARA81} - \c{ARA84}
in the meson cloud model (Fig.1) if positive pions in
the cloud prevail over negative ones ({\it i.e.} the contribution
of the diagram of Fig.1a is larger than the one of Fig.1b).

In this paper we want to pay attention to another interesting
problem, the unusual small-$x$ behaviour of the spin structure
function $g_1$ of proton and neutron \c{EMC}--\c{HERMES}. The
situation is especially evident in the case of the neutron
structure function $g_1^n$, which is very small in the
region of $x\gtrsim 0.2$ and  goes to rather large negative
values as $x$ decreases to $x\sim 10^{-2}$ ~\cite{E154}. This
behaviour cannot be explained by valence quark models, first
since valence quarks contribute to the function $g_1$ mainly at
$x\gtrsim 0.1$ and second, because of strong cancellation of
valence quark contribution for the neutron~\footnote
{Thus the neutron is a good laboratory for the investigation of
non-valence components of the nucleon.}.
At the same time the observed small-$x$ behaviour contradicts to
the one expected for the $A_1$ Regge-pole exchange because of its
low intercept. It would be natural to attribute this to the
manifestation of the nucleon sea. However, it is usually believed
that the meson cloud model cannot lead to a polarization of sea
quarks~\cite{ZOL92}, as the diagrams of Fig.1 with $\pi\pi$
exchange in the $t$ channel do not contribute to the function
$g_1$. In this paper we would like to emphasize that an
interference between $\pi$ and $\rho$ exchanges ($\pi\rho$
exchange in the $t$-channel -- Fig.2) leads to a substantial
polarization of the nucleon structure function $g_1$ with isospin
$I=1$ in the $t$-channel. It will be shown also that flavour
asymmetry of the polarized quark sea must be even stronger than
for the unpolarized case. Experimental tests of this model are
proposed.
\section*{2. General analysis of the small-$x$ behaviour of the
spin--dependent structure functions}
\indent

Analysis of the small-$x$ behaviour of structure functions of
deep--inelastic scattering on a nucleon in terms of the leading
$j$-plane singularities has been performed many years ago
\c{HEI73}. However there is a controversy concerning a
possibility of different contributions to functions $g_1(x)$,
$g_2(x)$ \c{CLO94}. So we shall repeat here the main consequences
of such an analysis.

The polarized structure functions $g_1(x,Q^2)$ and $g_2(x,Q^2)$
define the antisymmetrical in $\mu ,~\nu$ part of the amplitude
of the virtual Compton scattering (see {\it e.g.} \c{IOF84},
\c{ANS95}): \begin{equation} \label{W} W_{\mu\nu}^{(A)}(q;P,S)=
{2m_N\over (P\cdot q)}
\epsilon_{\mu\nu\rho\sigma}q^{\rho}
\left[ g_1(x,Q^2)S^{\sigma}+g_2(x,Q^2)S_{\perp}^{\sigma}\right] ~~,
\end{equation}
where $q$ is a photon 4-momentum, $Q^2=-q^2$, $P$ and $S$ are
nucleon momentum and spin vectors, and
\begin{equation}
S_{\perp}^{\sigma}=S^{\sigma}-{(S\cdot q)P^{\sigma}\over
(P\cdot q)} ~~.
\end{equation}

The absorptive parts of the $s$-channel helicity amplitudes
$F_{\lambda_{\gamma^{\ast}}^{'}\lambda_{N}^{'};
\lambda_{\gamma^{\ast}}\lambda_{N}}$
for forward $\gamma^{\ast}N$ scattering can be written in terms
of the functions $g_1$ and $g_2$ as follows \c{HEI73}
\bal
\label{comb1}
&& {1\over 2}\left(F_{1\pm{1\over 2};1\pm{1\over 2}}-
F_{-1\pm{1\over 2};-1\pm{1\over 2}}\right) =
\pm{1\over m_N}\left( g_1(x,Q^2)-{4m_N^2\over Q^2}x^2 g_2(x,Q^2)\right)
~,   \\
\label{comb2}
&& F_{0\mp{1\over 2};\pm 1\pm{1\over 2}}=
{\sqrt{2Q^2}\over m_N\nu} \left(g_1(x,Q^2)+ g_2(x,Q^2)\right) ~~.
\eal
The rules for construction of helicity amplitudes with definite
$t$-channel quantum numbers are well known (see e.g. \c{HEI73}).
In the combination of Eq.(\ref{comb1}) of $s$-channel helicity
amplitudes only states with $\sigma P=-1$ and $G(-1)^I\sigma=-1$
contribute asymptotically as $x\rightarrow 0$ (here $\sigma$ is
the signature, $P$ -- parity, $G$ -- $G$-parity and $I$ is the
isospin). These are the singularities (e.g. Regge poles) of the
so called axial group. Among known Regge poles the reggeons $A_1$
($I=1$) and $f_1$ ($I=0$) belong to this group.

The amplitude (\ref{comb2}) corresponds to the states in the
$t$-channel, which do not have definite parity (for example Regge
-cuts). The rightmost singularity of this type will be a cut due
to exchange by several Pomerons
\footnote { ~Numerically this
contribution is probably very small, as it is necessary to have
at least three Pomerons, and it is proportional to small
spin-flip couplings of the Pomeron.}.

Let us note that Regge cuts should also obey the quantum number
selection rules. In particular, Regge cuts due to exchange by any
number of Pomerons in the $t$-channel do not contribute (in the
leading order in $1/x$) to the function $g_1$. This is connected
to the fact that signature of the cut is equal to a product of
signatures of exchanged reggeons \c{GRI64} and is $+1$ for any
number of exchanged Pomerons. So the product $G(-1)^I\sigma$ in
this case is $+1$ and does not correspond to the amplitude $g_1$
we are interested in. This result can be generalized to the cuts
due to exchange of any number of reggeons with quantum numbers
$\sigma P=+1$, $G(-1)^I\sigma=+1$ ($P,f,\omega,\rho,A_2,\ldots$)
or $\sigma P=-1$, $G(-1)^I\sigma=+1$ ($\pi,\eta,\ldots$) if not
more than one of them has $I=1$. For any pair of such reggeons
$\sigma=\sigma_1\cdot\sigma_2$, $I=I_1+I_2$, $G=G_1\cdot G_2$ and
$G(-1)^I\sigma=G_1(-1)^{I_1}\sigma_1\cdot G_2(-1)^{I_2}\sigma_2
=+1$. A situation is different if both reggeons have $I_i=1$ and
add to the total isospin $I=1$. In this case $I=I_1 +I_2 -1$ and
the product $G(-1)^I\sigma=-1$. So the leading cuts contributions
to the small-$x$ behaviour of $g_1$ is given by $\rho A_2$,
$\rho\pi$ ($I=1$, $G=-1$) cuts. In the following we will estimate
these contributions in a simple model of a meson cloud and will
show that these contributions (especially the $\pi\rho$ one) can
be important in the preasymptotic region $10^{-2}\le x\le 0.1$.

Let us emphasize here that this classification of leading $j$-plane
singularities is valid in the leading (in $1/x$) approximation and
contributions damped by extra $x$-factor do not satisfy to these selection
rules. For example gluonic ladders can contribute to $g_1$ with $I=0$ at small
$x$. In the following we will consider only the leading contributions for
the $I=1$.
\section*{3. The meson cloud model}
\indent

It is known for a long time that $\pi$ exchange plays an
important role in the dynamics of peripheral inelastic
interactions of hadrons (for review see, for example, refs.
\c{BOR73} - \c{SPE98}). In a series of papers \c{BOR73},
\c{PON76}, \c{BOR78}, \c{ARA81}, \c{ARA84} the model with
"effective" reggeized pion exchange, which takes into account a
contribution of other exchanges ($\rho ,A_2,\: \ldots$), has been
developed and applied to a broad class of exclusive and inclusive
processes in hadronic interactions. It was demonstrated that the
model reproduces many characteristic features of these processes
over a wide region of energy. We mention here two qualitative
effects which are relevant to this subject.

It has been shown \c{BOR78} that the model leads to the
nontrivial behaviour of spin-dependent residues for the leading
Regge poles, -- small flip-nonflip ratio for $I=0$ reggeons
($P,f$) and large one for $I=1$ exchange ($\rho$ reggeon), in an
agreement with experimental observations. Note that the specific
spin and isospin structure of the bottom blob in Fig.1 due to
pion quantum numbers plays a crucial role in description of
observed spin structure. There is a strong cancellations between
contributions of the nucleon and $\Delta$ isobar intermediate
states for particular amplitudes.

Another example is related to the flavour structure of the quark
sea in the nucleon mentioned in Introduction. It was first noted
by Sullivan \c{SUL72} that $\pi$ exchange can contribute also to
the deep inelastic scattering (DIS) (Fig.1). In refs. \c{ARA81},
\c{ARA84} the effective pion exchange model has been applied to
calculation of antiquark distributions in a nucleon, using the
distribution of $\bar{q}$ in the pion. It was noted (see also
ref. \c{THO83}) that in this model $\bar{d}$
and $\bar{u}$ distributions in a nucleon are generally different.
This difference is connected with the relative contributions of
nucleon and $\Delta$ isobar in an intermediate state.

The distribution of antiquarks in the nucleon in this model has a
simple probabilistic interpretation and can be written as a
convolution of the probability $w_{\pi^k/N}$ to find a pion
$\pi^k$ in a nucleon $N$ with the probability $\bar{q}_{\pi^k}$
to find an antiquark in the pion, summed over all types
of pions \c{ARA81}, \c{ARA84}:
\begin{equation}
\label{qpipi}
\bar{q}_N(x)=\sum_k\int_x^1 {dx_{\pi}\over x_{\pi}}\:
w_{\pi^k/N}(x_{\pi})\:\bar{q}_{\pi^k}(x/x_{\pi}) ~~.
\end{equation}
The probability $w_{\pi^k/N}(x_{\pi})$ is a sum of probabilities
$w_k^{(R)}(x_{\pi})$ of production of a given state
$R=N,\Delta$. The latter ones expressed in terms of known
$\pi NN$, $\pi N\Delta$ couplings and form factors
$G_{\pi}^{R}(t)$ which determine pion off-shell dependences
\begin{equation}\label{wpipi}
w_k^{(R)}(x_{\pi})={x_{\pi}\over 16\pi^2}
\int_{-\infty}^{\tau_R(x_{\pi})}
g_{\pi^k NR}^2 (t)\left( G_{\pi}^{R}(t)\right)^2 ~~, \quad (R=N,\Delta ) ~~.
\end{equation}
Here $\tau_R$ is a minimal momentum transfer to the baryon $R$,
\begin{equation}
\tau_R(x_{\pi})\approx -{m_R^2 x_{\pi}\over 1-x_{\pi}}+m_N^2x_{\pi} ~~,
\end{equation}
and $t$ is connected with $\tau_R$ and transverse momentum of
produced baryon $\bf{k_{\perp}}$ as
\begin{equation}
t\approx\tau_R-{{\bf k}_{\perp}^2\over 1-x_\pi} ~.
\end{equation}
The functions $g_{\pi^k NR}(t)$ contain beside corresponding
coupling constants also traces over baryon indices providing
extra $t$ dependence of the bottom blob.

These formulas correspond to the simple pion cloud model, --
there is a cloud of pions in a fast moving nucleon and a virtual
photon interacts with an antiquark (quark) of the pion. In this
formulation the pion is an "effective" one and other exchanges
are taken into account by a proper choice of "form factors"
$G_R(t)$. A formulation of the meson cloud model in terms of
light-cone wave functions of a pions (and other pseudoscalar and
vector mesons) in a nucleon
\footnote{
It is worth to emphasize at this place that in contrast to
the light-cone approach we do not consider mesonic cloud in
terms of relativistic wave function. It means that interactions
with the meson and the "bare" nucleon (or $\Delta$-isobar) in
Figs.3a,b are not considered as interactions with the two
components of the physical nucleon. We treat these diagrams as
Reggeon exchanges, and the main preference of the pion exchange
is connected to the smallness of the pion mass and therefore to
the close position of the corresponding pole to the physical
region. The $t$-dependent residue of the Regge-pole amplitude can
be interpreted as a "form factor" function describing off-shell
behaviour of pionic amplitudes. In the relativistic wave-function
approach mesonic couplings are described in terms of
$x,p_{\perp}$ dependence of the wave function, and as a result
one has exponential damping for $x_{\pi}\rightarrow 0$ where pion
exchange dominance is well justified and confirmed by
experiments.}
was used by several authors
\c{HEN90}-\c{JEN91}, \c{ZOL92}, \c{MUL92}-\c{HOL96}, \c{SPE98}
and has been applied to different aspects of DIS, including the
"spin crisis" \c{ZOL92}, \c{HOL96}. However the main attention
was paid to a calculation of the function $g_1(x,Q^2)$ resulting
not from the meson cloud but from the "core" baryon, because the
diagrams of Fig.1 do not contribute to $g_1(x,Q^2)$. It has been
shown \c{ZOL92} that the existence of the meson cloud leads to a
substantional modification of the polarized quark contributions
in the region $x\sim 1$ (mainly due to renormalization of the
nucleon wave function), but the interaction with the core baryon
(the diagrams of Fig.3b) does not influence the polarization of
antiquarks from the nucleon sea.

The experience of dealing with the pion exchange model has shown
that, if the pion exchange is not forbidden for a particular
process, it dominates. It is related to the smallness of the pionic
mass and therefore to the close position of the corresponding pole
to the physical region of the process. Unluckily, as it was shown
in section 2, the $\pi\pi$ exchange does not contribute to the
amplitude connected to the $g_1$ structure function. Here we
would like to consider a new contribution of the diagram of
Fig.2. It contains $\pi\rho$ exchange in the $t$-channel and
corresponds to an interference between $\pi$ and $\rho$
exchanges\footnote{~Note that since our parameterization of the pion
exchange contains partly the $A_2$ contribution we take into account
effectively also the $A_2-\rho$ cut.}.
According to a general classification given in section 2
such a state can contribute to the small -x limit of function $g_1$
with the total isospin $I=1$ in the $t$-channel. This means that its
contributions to $g_1$ of proton and neutron have different
signs. The new element of this model is a strong polarization of
a nucleon sea and, in particular, of antiquarks of the nucleon. We
will show below that this polarization is strongly flavour
dependent.

Let us note here that the model gives a nonperturbative input for initial
condition of QCD-evolution. So its predictions should be valid at 
values of $Q^2 \sim 1~GeV^2$ and will be modified by perturbative effects
at much larger $Q^2$.

\section*{4. Description of the model}
\indent

Let us consider the diagrams of Fig.4 to estimate the
$\pi\rho$-exchange contribution to the $g_1$ structure function.

As we are interested in the imaginary part of the amplitude,
particles in the intermediate state are on the mass shell. In order
to get the partonic interpretation for the contribution of these
diagrams it is convenient to use the light-cone variables for
momenta in the top part of the diagram and to choose a Lorentz
frame with zero 'plus'-component of the photon momentum,
$q^+\equiv q_0+q_z=0$. This choice allows us to avoid non-partonic
contributions \c{MIC82}. In the Lorentz frame, where the proton
is moving along the $z$ axis, one has simple kinematic relations
$$
S_{\perp}^+=0 ~~,\quad 2(P\cdot q)=P^+ q^- ~~,
$$
and the tensor $W_{\mu\nu}$ from Eq.(\ref{W}) is determined at
$\mu =x$ and $\nu =y$ by only the $g_1$ structure function:
\begin{equation}
W_{xy}^{(A)}=4g_1(x,Q^2) ~,\quad({\rm for~ }q_+=0) ~.
\end{equation}

A standard consideration of the upper blob (see {\it e.g.}
\c{ANS95}) reduces the answer after antisymmetrization in
the photon polarization indices $\mu,\nu$ to the calculation of
the matrix element for the axial current component
$A^+=\gamma_5\gamma_+$ between pion and $\rho$ meson states. The
off-shell part of the quark propagator (the 'instantaneous'
component \c{MIC82}) is proportional to $\gamma^+ = \gamma_0
+\gamma_3$ and, since $\gamma_+^2 =0$, it is possible for the
'plus'-component of the current to consider the struck quark as
being on the mass shell as it is supposed in the partonic
picture. This prescription corresponds to the light-front
formalism \c{MIC82}.

Since axial curent $A^+$ measures the helicity of the (on-shell)
quark or antiquark:
\begin{equation}
\label{hel}
\bar{u}_{p,\lambda}\gamma^{+}\gamma_5 u_{p,\lambda}=
-\bar{v}_{p,\lambda}\gamma^{+}\gamma_5 v_{p,\lambda}=
4\lambda p^+ ~~,
\end{equation}
the result is expressed through the difference of contributions
for quarks of opposite helicities, i.e. doubled contribution to
the $\pi\rho$-interference term from the quark with positive
helicity as it is shown schematically in Fig.5.

For the whole diagram one gets the well known relation, which
connects the $g_1(x)$ structure function and the difference of
the polarized quark distributions $\Delta q_N (x)$:
\bal
\label{g1}
&& g_1(x)={1\over 2}\sum_q e_q^2 \Delta q_N(x) ~~, \nn \\
&& \Delta q_N(x)= q_N^+(x)-q_N^-(x)-\bar{q}_N^+(x)+\bar{q}_N^-(x)
 ~~,
\eal
and the part of the difference due to $\pi\rho$ interference has
the form similar to Eq.(\ref{qpipi}) :
\begin{equation}
\label{qpirho}
\Delta q_N (x)=\sum_k\int_x^1 {dx_{\pi}\over x_{\pi}}\:
w_{\pi\rho}^{\alpha}(x_{\pi})\:
\Delta q_{\pi\rho}^{\alpha}(x/x_{\pi}) ~~.
\end{equation}
Here $\alpha$ is a Lorentz vector index of the exchanged
$\rho$--meson (we omit indices specifying charges of $\pi ,\rho$ 
mesons), $w_{\pi\rho}^{\alpha}$ includes in analogy with
Eq.(\ref{wpipi}) the contribution $B_{\pi\rho}^{\alpha}(t)$ from
the bottom (nucleon) blob of the diagram of Fig.4 for positive
helicity of the initial nucleon and meson "form factors"
$G_{\pi ,\rho}^{R}(t)$ (propagators and off-shell factors):
\begin{equation}
\label{wpirho}
w_{\pi\rho}^{\alpha}(x_{\pi})={x_{\pi}\over 16\pi^2}
\sum_{R=N,\Delta}\int_{-\infty}^{\tau_R(x_{\pi})}
B_{\pi\rho}^{\alpha}(t)\: G_{\pi}^{R}(t)\: G_{\rho}^{R}(t) ~~,
\end{equation}
and $\Delta q_{\pi\rho}^{\alpha}$ is a difference of polarized
quark contributions coming from the top (quark) blob.

Let us discuss now separately the main ingredients entering
Eqs.(\ref{qpirho}), (\ref{wpirho}).

\noindent
\underline{{\em (i) The quark blob, $\Delta q_{\pi\rho}^{\alpha}$}}
\nopagebreak

It is convenient to write down the part coming from the upper
blob in terms of internal quark variables of the meson state
$x_q, {\bf q}_{\perp}$, which are related to the variables $x,
{\bf p}_{\perp}$ of the struck quark through the relations
\begin{equation}
x_q=x/x_{\pi} ~~, \quad
{\bf p}_{1\perp}=x{\bf k}_{\perp}+{\bf q}_{\perp} ~~.
\end{equation}

It will be assumed in the following that the {\em nonpolarized}
quark distributions are the same for the $\pi$ and $\rho$ mesons:
\begin{equation}
f_{q/\pi}(x_q,{\bf p_{qT}})=f_{q/\rho}(x_q,{\bf p_{qT}})=
f_q(x_q,{\bf p_{qT}}) ~~,
\end{equation}
with the normalization condition
\begin{equation}
\int_0^1 dx_q d^2{\bf q}_{\perp} f_q(x_q) =1 ~~.
\end{equation}
Then
\begin{equation}
\Delta q_{\pi\rho}^{\alpha}(x_q)= (N_{\pi}N_{\rho})^{-1/2}
\int d^2 q_{\perp}
f_q(x_q,{\bf p_{qT}}) N^{\alpha} ~~.
\end{equation}
Here $N_{\pi},N_{\rho}$ are normalization functions and the function
$N^{\alpha}$ contains spin-dependent contribution corresponding
to the non-diagonal $\pi\rho$ and $\rho\pi$ transition:
\be
N^{\alpha}=N_{\pi\rho}^{\alpha}+N_{\rho\pi}^{\alpha} ~~,
\ee
\be
N_{\pi\rho}^{\alpha}=\Tr\left[
(\hat{p}_1+m_q){1-\gamma_5\hat{s}_1\over 2}\Gamma^{(\pi)}
(-\hat{p}_2+m_q)\Gamma_{\alpha}^{(\rho)}
\right] ~~,
\ee
($s_1$ is a spin vector of the quark struck by the photon), and
similarly (with interchange of $\pi$ and $\rho$ vertices) for
$N_{\rho\pi}^{\alpha}$.

It is important that $p_1$ and $p_2$ in the equation above are
on-mass-shell 4-momenta of the quarks, $p_1^2=p_2^2=m_q^2$, and
the 'minus' component of momentum is not conserved in the
meson--quark--quark vertices: $p_1^-+p_2^-\ne k^-$. (These
prescriptions have been accepted just at the point where
relations (\ref{hel}) have been used). As a result the following
kinematic relation is valid:
\be
(p_1 + p_2)^2=M_0^2={{\bf p}_{qT}^2+m_q^2\over x(1-x)} ~~,
\ee
where $M_0$ is a standard variable in the light-cone formalism.

The spin structure of the meson--quark vertices has been chosen as
\begin{equation}
\Gamma^{(\pi)}=\gamma_5 ~~, \quad
 \Gamma_{\alpha}^{(\rho)}=\gamma_{\alpha} ~~.
\end{equation}
The normalization functions $N_{\pi}, N_{\rho}$ can be obtained {\it
e.g.} by consideration of the diagonal electromagnetic form factors
at zero momentum transfer. This gives
\bal
& N_{\pi}=\Tr \left[
(\hat{p_1}+m_q)\gamma_5 (-\hat{p_2}+m_q)\gamma_5\right]
~, \\
& N_{\rho}={1\over 3}\Tr \left[
(\hat{p_1}+m_q)\gamma_{\alpha}(-\hat{p_2}+m_q)\gamma_{\alpha}
\right]
~,
\eal
for the diagonal $\pi\pi$ and $\rho\rho$ transitions
correspondingly. The quark mass $m_q$ was taken as $m_q=0.3~GeV$
for calculations. It was pointed out above that we estimate the large
distance contribution of the meson cloud to the function $g_1$,
so the mass of the quark in our model should be considered as a
constutuent quark mass and we take for it the value $m_q=0.3~ GeV$.

To show an important role of the choice of the distribution
$f_q(x_q)$ we tried two extreme cases -- a simple wave function
description which assumes quark--antiquark content of the pion
\c{RWF} and the Gl\"{u}ck--Reya--Stratmann parameterization \c{GRS}
which has a singularity at $x=0$. The parameterizations used for
calculations are listed in the Appendix.
\\ \noindent
\underline{{\em (ii) The nucleon blob, $B_{\alpha}$}}

The bottom blob with $R=N,\Delta$ in the intermediate state
contains the nucleon traces
\begin{equation}
B^{\alpha}=\sum_R (B_{\pi\rho}^{(R)\alpha}+B_{\rho\pi}^{(R)\alpha}) ~~,
\end{equation}
where
\bea
B_{\pi\rho}^{(N)\alpha}=
\Tr\left[{(\hat P}+m_N){1+{\hat S}\gamma_5 \over 2} \Gamma_{\pi NN}
D^{(N)}(P-K)\Gamma_{\rho NN}^{\alpha} \right] \\
B_{\pi\rho}^{(\Delta)\alpha\tau}=
\Tr\left[{(\hat P}+m_N){1+{\hat S}\gamma_5 \over 2}
\Gamma_{\pi N\Delta}^{\tau}
D_{\tau\sigma}^{(\Delta)}(P-K)\Gamma_{\rho NN}^{\alpha\sigma}
\right]
\eea
The structure of the vertices $\Gamma_{\pi NR}$ and $\Gamma_{\rho
NR}$ and propagators $D^{(R)}$ is given in Appendix.

Let us note that for {\em unpolarized} quark sea the contribution to
the bottom blob due to nucleon and $\Delta$ isobar in the
intermediate state are comparable in the magnitude. The
particular values of coupling constants lead to larger values for
the nucleon contribution and as a result the flavour asymmetry of
antiquark sea of the proton in favour of $\bar{d}$ antiquarks. It
can be shown that for the {\em polarized} quarks the contribution
of $\Delta$ isobar is strongly suppressed due to spin structure
of vertices. Moreover, due to $C$-parity limitation only
exchanges of charged  $\pi,\rho$ mesons are allowed. It results
in much more drastic flavour asymmetry for polarized quark sea
compared to the case of unpolarized sea.
\\ \noindent
\underline{{\em (iii) The form factors, $G_{\pi ,\rho}^R (t)$}}

We used the same $\pi$ meson form factor $G_{\pi}^R (t)$ as in
refs.\c{ARA81},\c{ARA84} which allowed to give a consistent
description of nucleon and $\Delta$ inclusive spectra, data on
Drell--Yan process, and some other inclusive and exclusive
processes in the framework of the pion exchange model. The
function $G_{\rho}^R$ describing the off-shell behaviour of the
$\rho$ meson propagator is known much worse
\footnote{~We have not used the reggeized version for the off-shell
behaviour of the $\rho$ exchange as it is difficult to find
one which would be valid also for not small values of $x_{\rho}$.}.
We used the simple
exponential damping. The formulas for the form factors
are given in the Appendix.
\section*{5. Comparison with experiment}
\indent

The model formulated above allows us to calculate sea-quark
polarization to the structure function $g_1$ of the nucleon with $I=1$.
We have not tried to calculate a nonperturbative contribution to the
function $g_1$ with $I=0$ (for calculation of $\rho- \rho$ exchange to this
function see ref.\c{FRI98}). Experimental data show that at small
$x\sim 10^{-2}$ the amplitude with $I=1$ in the $t$-channel gives the
main contribution.  In the following we will compare our results with
the $g_1$ of the neutron, because in this case a contribution of the
valence quarks is small and contribution
of the sea can be more easily separated compared to the proton case.
The results for the $g_1^p$ are just opposite in sign due to $I=1$ of
the considered contribution. Note that gluons (and in particular
the gluon anomaly \c{ANS95}) do not contribute to this isospin
combination. These contributions can be important for $I=0$ amplitude.

Our model depends on several parameters which are not quite well known 
(e.g. $g_{\rho NN}$ coupling constant) and on assumptions listed above.
So we do not pretend for detailed descriptions of the I=1 sea component 
of the structure function $g1$ but rather for a qualitative effect and 
demonstrate that this effect is of right order of magnitude.

In Fig.6 the contribution of the simplest diagrams of Fig.2
with only $q\bar{q}$ component of the pion ($\rho$-meson) wave function
is shown. Due to a fast decrease of the wave functions as
$x_q\rightarrow 0$ $g_1$ decreases for $x<5\cdot 10^{-2}$. It is
known, however, that the valence quark distributions in the pion
are substantially different (especially in the region of small
$x_q$) from the behaviour corresponding to the simplest
$q\bar{q}$ wave function. For a distribution of valence quarks
given by GRS model \c{GRS} we obtain theoretical curves shown in
Fig.7, which are in a good agreement with experimental data. Thus
the model gives a reasonable estimate for both the size and the
form of the sea-quark polarization in the nucleon. A word of
caution concerning the above estimate of the valence quarks in
the pion. The small-$x$ behaviour of the valence quarks of the pion in the
GRS model $\sim 1/x^{\gamma}$ with $\gamma\approx 0.5$ is
strictly speaking true for a "diagonal" transition with quantum
numbers of $A_2,f$ in the $t$-channel. In our case we should
know the behaviour of the analogous quantity with axial quantum
numbers in the $t$-channel. In the model \c{GRI82}, where
corresponding reggeons are generated due to gluon emissions in
the box diagram (Fig.8) it is possible to show that a difference
between vector and axial vector contributions appears only at
three-loop level (emission of two gluons in the $s$-channel of
Fig.8). So this difference should be important in the upper blob
only at small $x_q<x_q^{cr}\approx 0.1$. This means that our
estimate with GRV structure function of the pion can be valid up
to $x\sim \bar{x}_{\pi}x_q^{cr}\sim 0.01$. At smaller values of
$x$ we expect a gradual transition to the behaviour corresponding
to the $A_1$ Regge exchange.

We would like to emphasize that due to a smallness of the diagram
with the $\Delta$ intermediate state of Fig.2 only the $\bar{d}$
quarks of the proton ($\bar{u}$ of the neutron) are polarized.
The following relations for polarized quark distributions in the
proton and the neutron are valid in the model (here we separate 
quark and antiquark polarized distributions unlike the notation 
of Eq.(\ref{g1})):
\bea
\Delta u_{sea}^p (x)=\Delta d_{sea}^n (x)
=\Delta {\bar d}^p(x)=\Delta {\bar u}^n(x)\approx
6 g_1^p (x)=-6g_1^n (x) ~~, \\
\Delta d_{sea}^p (x)=\Delta u_{sea}^n (x)
=\Delta {\bar u}^p (x)=\Delta {\bar d}^n (x)\approx 0 ~~.
\eea
Thus the model predicts even higher flavour asymmetry for
polarized antiquarks than for unpolarized distributions. Present
estimate of polarization for antiquarks do not contradict to
limits on this quantity established from a study of semiinclusive
DIS \c{SMCsemi98}, especially taking into account that the
hypothesis of flavour symmetry for light quarks (antiquarks) of
the sea has been used in the analysis of experimental data.
In principle, different flavour contributions to the polarized quark
sea can be separated and model predictions can be tested if accuracy
of experimental data will be improved.

The most straightforward test of our model would be a
measurement of the polarization asymmetries in DIS with
simultaneous registration of a nucleon or $\Delta$ in the final
state. For $\Delta$ production sea polarization should be
absent (or small) compared to the case of neutron production
off proton (or proton from a neutron target).

Let us comment on a relevance of these results for the Bjorken
\c{BJO66} and Ellis--Jaffe \c{ELL74} sum rules. It is clear from
experimental results \c{E143} that the region of $x<0.1$, where
sea quarks are important gives a substantional contribution
($\sim 30 \%$)
to the Bjorken sum rule. This means that
sea-quark component of a nucleon should be important in
dynamical calculation of the axial coupling constant $g_A$. So
all calculations of the axial couplings of baryons \c{HOL96}
should be reconsidered with an account of this mechanism. Another
important problem for the test of the Bjorken sum rule is the
evaluation of the integral of the difference
$g_1^p(x)-g_1^n(x)$ in the unmeasured region of very small $x$
\c{E143}. The simplest power-law extrapolation of existing data
gives a large contribution to the integral \c{E143}. Our
investigation indicates that a steep increase of the sea
component with $I=1$ in the structure function $g_1(x)$ should change
at $x\sim 10^{-2}$ to a slower increase (or constant behaviour)
as $x$ decreases. In this case the unmeasured small-$x$ region
contributes only $0.02\div 0.03$ to the integral.

It is known \c{ZOL92}, \c{HOL96} that an inclusion of mesonic
cloud to the nucleon wave function leads to a better agreement
with Ellis--Jaffe sum rule. However it is difficult to obtain a
good description of experimental data on $g_1^p$, $g_1^n$ and to
solve completely the problem of "spin crisis". We hope that
situation will improve with an account of interference diagrams
considered here. An interference of $K$ and $K^{\ast}$ exchanges
will lead to some polarization of $s$ and $\bar{s}$ quarks, which
can be relevant for this problem.
\section*{6. Conclusions}
\indent

In this paper we proposed a new nonperturbative mechanism for
polarization of sea quarks in a nucleon. The mechanism leads to a
strong flavour dependence of the sea polarization. It can not be
imitated by the diagrams of perturbative QCD. An estimate of the
$\pi\rho$ interference mechanism carried out in this paper shows
that it can account for an unusual behaviour of the structure
function $g_1^n$ in the small-$x$ region and is important for
calculations of the axial couplings of baryons.

Future experiments with polarized beam and target in DIS and
hadronic interactions will allow a critical test of the proposed
mechanism for the sea-quark polarization.
\\[2mm]
{\small
{\bf Acknowledgments}

The authors thank B.L.~Ioffe and J.H.~Koch for helpful discussions.
K.B. wants to gratitude the Instituto de Ciencias Nucleares,
UNAM (Mexico) and especially A.~Turbiner for their kind hospitality
where the present work was finished.
This work was
supported in part by INTAS grant 93-0079ext, NATO grant OUTR.LG
971390, RFBR grants 96-02-191184a, 96-15-96740 and
98-02-17463 and DGAPA grant IN 105296.
}
\section*{Appendix}
\indent

Here we give a description of the main components entering the
model -- quark distributions used for calculations, baryon
vertices and propagators, and form factors.
\\[2mm]
{\bf A. Quark distributions in meson}
\setcounter{equation}{0}
\def\theequation{A.\arabic{equation}}

In order to illustrate an effect of the choice of different
parametrization  we used two types of distributions.
The first one supposes that mesons contain only valence
quark and antiquark. This two-particle distribution was
parametrized in terms of the light-cone wave function \c{RWF}:
\be
f_q(x_q,{\bf p}_{qT})=|\psi(x_q,{\bf p}_{qT})|^2 ~,
\ee
where
\be
\psi(x_q,{\bf p}_{qT})=\sqrt{M_0 \over 2x_q(1-x_q)}\phi(p)  ~~,
\ee
with
\bea
&&M_0^2={m_q^2+{\bf p}_{qT}^2 \over x_q(1-x_q)} ~~~,  \\[2mm]
&&p=\sqrt{{\bf p}_{qT}^2+p_z^2} ~, \quad p_z=M_0(x_q-1/2) ~.
\eea
The function $\psi(x,{\bf p}_{T})$ is normalized as
\be
{1\over 16\pi^3} \int dx \int d^2 p_T
\left|\Psi(x,{\bf p}_{T})\right|^2 =1 ~,
\ee
what corresponds to the standard normalization of the function
$\phi(p)$:
\be
{1\over (2\pi)^3} \int d^3p \left |\phi(p)\right |^2 =1 ~~.
\ee
We use a simple oscillator-type parametrization for the
wave function $\phi(p)$ \c{RWF}:
\be
\phi(p)=\pi^{-3/4}\beta^{-3/2}(2\pi)^{3/2}
\exp(- p^2/2\beta^2) ~
\ee
with $\beta=0.32$.
The second type of distribution written under an assumption of
multiparticle content of a meson was taken from the paper by
Gl\"{u}ck, Reya and Stratmann \c{GRS}:
\be
f_q(x_q,{\bf p}_{qT})=0.471 x_q^{-0.501} (1-x_q)^{0.367}(1+0.632 \sqrt{x_q})
\; {\lambda \over \pi} \exp(-\lambda {\bf p}_{qT}^2) ~,
\ee
where the value of $\lambda$ does not effect the final result and was chosen
$\lambda =1$.
\\[2mm]
{\bf B. The baryon vertices}
\setcounter{equation}{0}
\def\theequation{B.\arabic{equation}}

The baryon vertices were taken in the form
\bea
\label{pi NN}
&&\Gamma_{\pi NN}=g_{\pi NN}\gamma_5 ~~,  \\
\label{rho NN}
&&\Gamma_{\rho NN}^{\alpha}=g_{\rho NN}^{(0)}\gamma_{\alpha}+
{g_{\rho NN}^{(1)}\over 2m_N}\sigma_{\alpha\beta}k_{\beta}~~, \\
\label{pi N Delta}
&&\Gamma_{\pi N\Delta}^{\tau}=g_{\pi N\Delta} p_{N\tau} ~~,
\eea
\vspace*{-10mm}
\setcounter{equation}{0}
\def\theequation{B.4\alph{equation}}
\bea
\label{rho N Delta}
&&\Gamma_{\rho N\Delta}^{\alpha\tau}= g_{\rho N \Delta}
\gamma_5 (\hat{k} g_{\alpha\tau}-\gamma_{\alpha}k_{\tau}) ~,  \\
\label{rho N Delta a}
&&\Gamma_{\rho N\Delta}^{\alpha\tau}= g_{\rho N \Delta}
\epsilon^{\alpha\tau\delta\gamma}P_{\delta}p_{\Delta}^{\gamma} ~,
\eea
\setcounter{equation}{4}
\def\theequation{B.\arabic{equation}}
\noindent
where $\alpha$ and $\tau$ are Lorentz indices of the $\rho$ and
$\Delta$ correspondingly.
The form of the first three vertices is unambiguious, and for the
forth one we tried both the structure (\ref{rho N Delta}) used in
meson exchange models \c{MHE87} and the simplest structure
(\ref{rho N Delta a}) which is in an agreement with experimental
data on the process $\pi p\rightarrow \pi\Delta$ \c{KAI66}.

Only the second term in Eq.(\ref{rho NN}) does contribute to the
small-$x$ asymptotics of the $g_1(x)$. The intermediate state
with $\Delta$ isobar does not contribute to the asymptotics of
the $g_1(x)$ at all (for the both structures (\ref{rho N
Delta}),(\ref{rho N Delta a})). This fact results in important
predictions on flavour structure of the polarized quark sea (see
section 5).

The following values of coupling constants were taken
\c{ARA84}, \c{MHE87}:
\bea
&{g_{\pi^0 pp}^2\over 4\pi} &=~ 14.6 ~, \\
&{g_{\rho^0 pp}^{(0)2}\over 4\pi} &=~ 0.84 ~, \quad\quad
g_{\rho NN}^{(1)}/g_{\rho NN}^{(0)} ~=~ 6.1 ~, \\
&{g_{\pi^- p\Delta^{++}}^2\over 4\pi} &=~ 19 ~GeV^{-2} ~, \\
&{g_{\rho^- p \Delta^{++}}^2\over 4\pi} &=~ 34.7 ~GeV  ~.
\eea
{\bf C. Baryon propagators}
\setcounter{equation}{0}
\def\theequation{C.\arabic{equation}}

Propagators of the nucleon and the $\Delta$-isobar have standard
form of those for particles of spin 1/2 and 3/2:
\bea
&& D_N(P)= \hat{P}+m_N ~,\\
&& D_{\Delta}(P)=(\hat{P}+m_N)\bigg[
{2\over 3m_{\Delta}^2}P_{\alpha}P_{\tau}-g_{\alpha\tau}
+{1\over 3}\gamma_{\alpha}\gamma_{\tau}+
{1\over
3m_{\Delta}}(\gamma_{\alpha}P_{\tau}-\gamma_{\tau}P_{\alpha})
\bigg] ~.
\eea
{\bf D. Form factors} \setcounter{equation}{0}
\def\theequation{D.\arabic{equation}}

The functions $G_{\pi}^R (t)$ were taken in the same
parameterization as in ref. \c{ARA84}:
\bea
&&G_{\pi}^{R}(t)=\exp\left[\left
(R_{1R}^2+\alpha^{\prime}\ln1/x_{\pi}\right)
(t-\mu^2)\right] \times \nn \\[2mm]
&&\times
\left \{
\begin{array}{ll}
 \pi\alpha^{\prime}/
(2\sin[\pi\alpha_{\pi}^{\prime}(t-\mu^2)/2])~,
& |t|\le|T_R| \cr
 \exp[R_{2R}^2(t-T_R)]
\pi\alpha_{\pi}^{\prime}/
2\sin[\pi\alpha_{\pi}^{\prime}(T_R-\mu^2)/2])~,
& |t|>|T_R| \cr
\end{array}
\right.
\eea
with parameter values
\bea
R_{1N}=0.3 ~Gev^{-2} ,\quad &R_{2N}=2.0 ~Gev^{-2} ,\quad
&T_N=-0.4 ~Gev^{2} \\
R_{1\Delta}= 0.2 ~Gev^{-2} ,\quad
&R_{2\Delta}=0.74 ~Gev^{-2} ,\quad,
&T_\Delta=-0.7 ~Gev^{2}
\eea
The values of the parameters provide continuity of the
derivative of function at $t=T_R$. In fact, as it was
mentioned, the contribution of $\Delta$-isobar is negligible
and only the nucleon intermediate state contributes to the $g_1$
structure function.

The function $G_{\rho}^{N}(t)$ was taken in the simple form
consistent with meson exchange models
\be
G_{\rho}^{N}(t)=\frac{e^{R^2 (t-m_{\rho}^2)}}{t-m_{\rho}^2} ~,
\ee
and results of calculations with $R^2=0.7~GeV^{-2}$ and $R^2=1.0~GeV^{-2}$ were
shown as an illustration.


\newpage
\begin{thebibliography}{99}
\bibitem{NMC} P.~Amaudruz {\it et al},
{\it Phys. Rev. Lett.} {\bf 66} (1991) 2712;
M.~Arneodo {\it et al}, {\it Phys. Rev.} D {\bf 50} (1994) R1
\bibitem{DY} A.~Baldit {\it et al},
{\it Phys. Lett.} B {\bf 332} (1994) 244
\bibitem{E866} E.A.~Hawker {\it et al},
{\it Phys. Rev. Lett.} {\bf 80} (1998) 3715
\bibitem{ARA81} G.G.~Arakelyan, K.G.~Boreskov, A.B.~Kaidalov,
{\it Sov. J. Nucl. Phys.} {\bf 33} (1981) 247
\bibitem{THO83} A.W.~Thomas,
{\it Phys. Lett.} B {\bf 126} (1983) 97
\bibitem{ARA84} G.G.~Arakelyan, K.G.~Boreskov,
preprint ITEP-50-1984, 1984;
{\it Sov. J. Nucl. Phys.} {\bf 41} (1985) 267.
\bibitem{EMC} J.~Ashman {\it et al} (EMC),
{\it Phys. Rev. Lett.} B {\bf 206} (1986) 364;
{\it Nucl. Phys.} B {\bf 328} (1989) 1
\bibitem{SMC} D.~Adams {\it et al} (SMC),
{\it Phys. Lett.} B {\bf 329} (1994) 399; B {\bf 357} (1995) 248;
B {\bf 396} (1997) 338; {\it Phys. Rev.} D {\bf 56} (1997) 5330
\bibitem{E142} P.L.~Anthony et al. (E142),
{\it Phys. Rev. Lett.} {\bf 71} (1993) 959;
{\it Phys. Rev.} D {\bf 54} (1996) 6620
\bibitem{E143} K.~Abe {\it et al} (E143),
{\it Phys. Rev. Lett.} {\bf 74} (1995) 346; {\bf 75} (1995) 25;
{\it Phys. Lett.} B {\bf 364} (1995) 61; SLAC-PUB-7753 (1998)
\bibitem{E154} K.~Abe {\it et al} (E154),
{\it Phys. Rev. Lett.} {\bf 79}(1997) 26;
{\it Phys. Lett.} B {\bf 404} (1997) 377;
{\it Phys. Lett.} B {\bf 405} (1997) 180;
\bibitem{HERMES} K.~Ackerstaff {\it et al} (HERMES),
{\it Phys. Lett.} B {\bf 404} (1997) 383.
%
\bibitem{ZOL92} V.R.~Zoller,
{\it Zeit. Phys.} {\bf C53} (1992) 443; {\bf C60} (1993) 141
\bibitem{HEI73} R.L.~Heimann,
{\it Nucl. Phys.} B {\bf 64} (1973) 429
\bibitem{CLO94} F.E.~Close, R.G.~Roberts,
{\it Phys. Lett.} B {\bf 336} (1994) 257
\bibitem{IOF84} B.L.~Ioffe, V.A.~Khoze, L.N.~Lipatov,
"Hard Processes", North Holland, Amsterdam, 1984
\bibitem{ANS95} M.~Anselmino, A.~Efremov, E.~Leader,
{\it Phys. Rep.} {\bf 261} (1995) 1
\bibitem{GRI64} V.N.~Gribov,
Proc. of XII Intern. Conf. on High Energy Physics,
Dubna, 1964, p. 394
\bibitem{BOR73} K.G.~Boreskov, A.B.~Kaidalov, L.A.~Ponomarev,
preprint ITEP-92-1973 (1973), published in
{\it Proc. of Yerevan Physics School}, 1975.
\bibitem{PON76} L.A.~Ponomarev,
{\it Sov. J. Part. Nucl.} {\bf 7}, 70 (1976).
\bibitem{SPE98} J.~Speth, A.W.~Thomas,
{\it Advances in Nuclear Physics}, {\bf 24}, 83 (1998).
\bibitem{BOR78}
K.G.~Boreskov, A.A.~Grigoryan, A.B.~Kaidalov, I.I.~Levintov,
{\it Sov. J. Nucl. Phys.} {\bf 27} (1978) 432
%
\bibitem{SUL72} J.D.~Sullivan,
{\it Phys. Rev.} D {\bf 5} (1972) 1732
\bibitem{HEN90}
E.M.~Henley, G.A.~Miller,{\it Phys. Lett.} {\bf B251} (1990) 453
\bibitem{SIG91} A.~Signal, A.W.~Schreiber, A.W.~Thomas,
{\it Mod. Phys. Lett.} {\bf A6} (1991) 271;
W.~Melnitchouk, A.W.~Thomas, A.I.~Signal,
{\it Zeit. Phys.} {\bf 340} (1991) 85
\bibitem{KUM91} S.~Kumano,
{\it Phys. Rev.} {\bf D43} (1991) 59; 3067
\bibitem{HWA91} W.-Y.P.~Hwang, J.~Speth, G.E.~Brown,
{\it Zeit. Phys.} {\bf A339} (1991) 461
\bibitem{JEN91} E.~Jenkins, A.V.~Manohar,
{\it Phys. Lett.} {\bf B255} (1991) 558; {\bf B259} (1991) 353
\bibitem{MUL92} P.J.~Mulders, A.W.~Schreiber, H.~Meyer,
{\it Nucl. Phys.} {\bf A549} (1992) 498
\bibitem{SZC93} A.~Szczurek, H.~Holtmann,
{\it Acta Phys. Pol.} {\bf B24} (1993) 1833
\bibitem{HOL96} H.~Holtmann, A.~Szczurek, J.~Speth,
{\it Nucl. Phys.} {\bf A596} (1996) 631
\bibitem{MIC82} C.~Michael, F.P.~Payne, {\it Zeit. Phys.} {\bf
C12} (1982) 145
\bibitem{RWF} W.~Jaus,
{\it Phys. Rev.} {\bf D44} (1991) 2851
\bibitem{GRS} M.~Gl\"{u}ck, E.~Reya, M.~Stratmann,
{\it Zeit. Phys.} {\bf C53} (1992) 651
\bibitem{FRI98} R.J.~Fries, A.~Sch\"{a}fer,
{\it Phys. Lett.} {\bf B443} (1998) 40
\bibitem{GRI82} A.A.~Grigoryan, N.Ya.~Ivanov, A.B.~Kaidalov,
{\it Sov. J. Nucl. Phys.} {\bf 36} (1982) 867
\bibitem{SMCsemi98} B.~Adeva {\it et al} (SMC),
Phys.Lett. {\bf B420} (1998) 180
\bibitem{BJO66} J.B.~Bjorken,
{\it Phys. Rev.} {\bf 148} (1966) 1467; {\bf D1} (1970) 1376
\bibitem{ELL74} J.~Ellis, R.L.~Jaffe,
{\it Phys. Rev.} {\bf D9} (1974) 1444; Erratum {\bf D10} (1974) 1669
\bibitem{MHE87} R.~Machleidt, K.~Holinde, Ch.~Elster,
{\it Phys. Rep.} {\bf 149} (1987) 1
\bibitem{KAI66} A.B.~Kaidalov, V.I.~Zakharov,
Pis'ma v JETP {\bf 2} (1966) 192
\end{thebibliography}
\end{document}